\documentclass[aps,prd,twocolumn,superscriptaddress,preprintnumbers,floatfix,nofootinbib,notitlepage,showkeys,showpacs]{revtex4}

\usepackage{graphicx}
\usepackage{graphics,psfrag}
\usepackage{latexsym}
\usepackage{amsmath,amssymb}
\usepackage{amsfonts}
\usepackage{enumitem}

\newcommand{\ev}[1]{\left \langle #1 \right  \rangle}

\newcommand{\be}{\begin{equation}}
\newcommand{\ee}{\end{equation}}
\newcommand{\bea}{\begin{eqnarray}} 
\newcommand{\eea}{\end{eqnarray}}
\newcommand{\nn}{~\nonumber \\}

\newcommand{\csb}{$\rm{\chi SB}$ }

\begin{document}

\title{Ideal Walking Dynamics via a Gauged NJL Model}
\author{Jarno Rantaharju}
\email{jmr108@phy.duke.edu}
\affiliation{Department of Physics, Box 90305, Duke University, Durham, NC 27708, USA}
\author{Claudio Pica}
\email{pica@cp3-origins.net}
\affiliation{CP3 -Origins \& IMADA, University of Southern Denmark, Campusvej 55, 5230 Odense, Denmark}
\author{Francesco Sannino}
\email{sannino@cp3-origins.net}
\affiliation{CP3 -Origins  \& Danish IAS, University of Southern Denmark, Campusvej 55, 5230 Odense, Denmark}
\affiliation{Theoretical Physics Department, CERN, Geneva, Switzerland}

\begin{abstract}
According to the Ideal Walking Technicolor paradigm large mass anomalous dimensions arise in gauged Nambu--Jona-Lasinio (NJL) models when the four-fermion coupling is sufficiently strong to induce spontaneous symmetry breaking in an otherwise conformal gauge theory. We therefore study the $SU(2)$ gauged NJL model with two adjoint fermions using lattice simulations.
The model is in an infrared conformal phase at small NJL coupling while it displays a chirally broken phase at large NJL couplings.
In the infrared conformal phase we find that the mass anomalous dimension varies with the NJL coupling reaching $\gamma_m \sim 1$ close to the chiral symmetry breaking transition, de facto making the present model the first explicit realization of the Ideal Walking scenario.
\end{abstract}

\preprint{CERN-TH-2017-053}
\preprint{CP3-Origins-2017-013 DNRF90}


\maketitle

\section{Introduction}

In   technicolor and (fundamental) composite Higgs models, four fermion interactions naturally emerge near the electroweak scale when trying to endow the Standard Model fermions with a mass term \cite{Weinberg:1975gm,Susskind:1978ms,Kaplan:1983fs,Kaplan:1983sm}. They   appear as an effective description of a more fundamental high energy model\footnote{For an explicit fundamental fermionic realization in both the technicolor and composite Higgs case using chiral gauge theories see \cite{Cacciapaglia:2015yra}. We note that a novel microscopic realisation for partial compositeness has been put forward in \cite{Sannino:2016sfx}. The recent fully composite realisation  overcomes earlier bottlenecks for partial compositeness, when expected to emerge  from pure fermion realisations, such as the unlikely existence of anomalously large anomalous dimensions \cite{Pica:2016rmv} for the composite baryon of the theory, and the fact that no truly fully viable microscopic description exists \cite{Ferretti:2016upr}. Interestingly four fermion interactions naturally emerge around the electroweak scale also in the fully fledged microscopic construction introduced in \cite{Sannino:2016sfx}. }.

Four fermion interactions can play a dual purpose, in addition to fermion mass generation, they can dramatically change the dynamics of the new strongly interacting sector \cite{Fukano:2010yv} and improve on the original walking paradigm \cite{Holdom:1981rm} by greatly extending the number of potential relevant theories that can be used to break the electroweak symmetry dynamically and, last but not the least increase the anomalous dimension of the technifermion mass operator \cite{Fukano:2010yv}. We refer also to \cite{Appelquist:1988fm,Kondo:1988qd,Holdom:1995fu} for pointing out the importance of the effects of strong four-fermion interactions on the dynamics of gauge theories.

In particular, in technicolor models producing the correct mass for the top quark requires balancing flavor changing neutral currents and the quark mass term arising from the same high energy interaction. In purely fermionic models of single-scale fermion mass generation the flavor changing neutral currents are suppressed by the high energy scale and the mass term can be enhanced in a walking technicolor model with a large mass anomalous dimension \cite{Holdom:1981rm}.
A number of issues plague the Walking realization within a gauge theory with a given fermion representation.
The first issue is that the number of flavors cannot be modified continuously reducing substantially the number of theories that can be just below the near-conformal transition.
Secondly neither higher-order precise computations\footnote{The first use of rigorous computations to elucidate the conformal dynamic properties of physical quantities such as the S-parameter appeared in \cite{Sannino:2010ca}. } \cite{Pica:2010mt,Pica:2010xq,Ryttov:2010iz,Ryttov:2016ner} nor lattice results \cite{Pica:2017gcb} so far support large enough mass anomalous dimensions for theories within the conformal window of \cite{Sannino:2004qp,Dietrich:2006cm,Pica:2010mt,Pica:2010xq}.
 Lastly we are not guaranteed that the transition is continuous in the number of flavors \cite{Sannino:2012wy}.

 However a significant four fermion coupling can increase the mass anomalous dimension \cite{Yamawaki:1996vr,Fukano:2010yv} while allowing to get arbitrarily close to the lower boundary of the conformal window \cite{Fukano:2010yv}. Ideally, a walking technicolor model could be constructed by allowing a strong four fermion interaction to induce chiral symmetry breaking when the gauged theory in absence of the four-fermion interactions is infrared conformal\footnote{Mass generation has also been analysed in the context of extra dimensional set-ups (see for instance~\cite{Fitzpatrick:2007sa,Albrecht:2009xr}), which however cannot be considered on the same footing as fundamental theories~\cite{Parolini:2014rza}. Recently, analyses using crossing symmetry in conformal field theories have added extra constraints for the scalar operator with the lowest dimension \cite{Rattazzi:2008pe,Rychkov:2009ij}. However, the constraints are not generally applicable to the operator relevant for flavor as demonstrated in ~\cite{Antipin:2014mga}.}.

It is for the reasons above that we investigate the $SU(2)$ gauged NJL \cite{Nambu:1961fr} (gNJL) model with two flavors of fermions transforming in the adjoint representation using lattice simulations. It was first realised and predicted in \cite{Sannino:2004qp,Dietrich:2005jn,Dietrich:2006cm} that the theory, at zero NJL coupling, could display (near) conformality. These results triggered a number of important lattice studies at zero NJL coupling \cite{Catterall:2007yx,Hietanen:2008mr,DelDebbio:2008zf,Catterall:2008qk,Bursa:2009we,DelDebbio:2009fd,DeGrand:2011qd, DelDebbio:2010hx,DelDebbio:2010hu,Patella:2012da,
    Giedt:2012rj, Bergner:2015jdn,Rantaharju:2015yva,DelDebbio:2015byq,Rantaharju:2015cne,Bergner:2016hip} agreeing on the infrared conformality of the theory.
Therefore, the gauged NJL model has an infrared conformal phase at small NJL coupling.
It is natural to expect that at strong  NJL coupling the four fermion interaction induce chiral symmetry breaking.
A similar model with no infrared conformal phase has been studied in \cite{Catterall:2011ab,Catterall:2013koa}.

To be precise here we study a model in which the NJL operator breaks the flavor symmetry and cannot be induced by the gauge coupling.
An infrared fixed point (IRFP) exists at vanishing NJL coupling.
It is possible that this fixed point may extend into a line of IR fixed points parametrized by the NJL coupling, with varying anomalous dimensions.
In this case, the mass anomalous dimension is expected to increase with the NJL coupling \cite{Fukano:2010yv}.  We study this possibility by measuring the anomalous dimension at several values of the NJL coupling.
Although here we focus on the infrared dynamics of the theory there are a number of interesting UV possibilities that are outside the scope of this work\footnote{For example, according to recent results, calculable in a perturbative regime, gNJLmodels can be seen as a special case of gauge Yukawa models \cite{Krog:2015bca}. A gauge Yukawa model, even when manifestly perturbative, under certain conditions can be viewed as a composite theory \cite{Krog:2015bca,Bardeen:1989ds,Chivukula:1992pm,Bardeen:1993pj} and reduce to a gNJL model at a high energy scale.
Furthermore, when the gauge coupling runs sufficiently slowly, a gNJL model may be renormalizable with a non-trivial coupling. In \cite{Kondo:1992sq} the model was studied in the limit of standing gauge coupling and was found to be renormalizable with a non-trivial NJL coupling. Similar models were studied in \cite{Harada:1994wy} and \cite{Kubota:1999jf} and were found to be non-trivial with sufficiently slowly running gauge coupling.
If a non-perturbative ultraviolet fixed point emerges in an NJL theory, the absence of an ultraviolet cutoff renders the theory fundamental according to Wilson. Recently the first rigorous results in four dimensions for the (non) existence of fully interacting ultraviolet fixed point appeared in (supersymmetric) gauged Yukawa theories in \cite{Litim:2014uca,Litim:2015iea,Intriligator:2015xxa,Bajc:2016efj}.  These results led to the recent discovery of the first example of a calculable radiative symmetry breaking  mechanism for UV complete QFTs at low energies akin to the radiative symmetry breaking that occurs in the Supersymmetric Standard Model \cite{Abel:2017ujy}.}.
Here we consider the model as an effective theory defined with an ultraviolet cut-off at some energy scale much higher than the inverse lattice spacing and we will not address the issue of taking the continuum limit.

In section \ref{themodel} of this work we introduce the lattice model in detail and discuss its symmetry properties. In section \ref{phasediagram} we present a sketch of the phase diagram. In section \ref{chiralbreaking} we study the chiral symmetry breaking transition and find the critical NJL coupling. In section \ref{anomdim} we measure the mass anomalous dimension as a function of the NJL coupling in the infrared conformal phase. Finally we conclude in section \ref{conclusions}.
We find numerical evidence to support the Ideal Walking scenario with mass anomalous dimensions growing towards unity as function of the NJL coupling from its infrared conformal value for the gauge theory.
Since the global symmetry of the model in the continuum, is U(1)$\times$ U(1) that breaks spontaneously to U(1) for large four-fermion coupling, the model can be viewed as a first step towards a model of dynamical electroweak symmetry breaking, but it can also be used in other model building contexts, including composite dark matter models and inflation \cite{Channuie:2011rq,Bezrukov:2011mv,Anguelova:2014dza,Anguelova:2015dgt,Channuie:2016iyy,Inagaki:2016vkf}.

\section{The Model}\label{themodel}

We study the $SU(2)$ gauged NJL model with 2 Dirac fermion flavors in the adjoint representation of the gauge group. The lattice action of the model is
\begin{align}
 S &= \beta \sum_{x,\mu<\nu} L_{x,\mu\nu}(U) \nn
 &+ \sum_{x,y} \bar\Psi(x) D_W(x,y) \Psi(y) + \sum_{x} m_0 \bar\Psi(x) \Psi(x)  \nn
  &- \sum_{x} a^2 g^2 \left [ \bar\Psi(x)\Psi(x)\bar\Psi(x)\Psi(x) \right ] \nn
  &- \sum_{x} a^2 g^2 \left [ \bar\Psi(x) i\gamma_5\tau^3\Psi(x)\bar\Psi(x) i\gamma_5 \tau^3\Psi(x) \right ], \label{action}
\end{align}
where $L_{x,\mu\nu}(U)$ is the Wilson plaquette gauge action and $U$ the gauge field, $D_W$ is the Wilson Dirac operator and $a$ is the lattice spacing. We perform lattice simulations using the Hybrid MonteCarlo (HMC) algorithm and handle the four fermion term using auxiliary fields:
\begin{align}
 S &=  \beta_L \sum_{x,\mu<\nu} L_{x,\mu\nu}(U) \nn
 &+ \sum_{x} \bar\Psi(x) \left [ D_W + m_0+\sigma(x) + \pi_3(x) i \gamma_5\tau^3 \right ] \Psi(x) \nn
 &+ \frac{\sigma(x)^2+\pi_3(x)^2}{4a^2g^2}. \label{actionaux}
\end{align}
The original action is recovered by integrating over the fields $\sigma$ and $\pi$.

The four fermion term preserves a $U(1)\times U(1)$ component of the full $SU(4)$ flavor symmetry of the gauge model. We will refer to the unbroken direction in the symmetry group as diagonal and the other directions as non-diagonal. While a $SU(2)\times SU(2)$ symmetry would be ideal, the auxiliary field representation would suffer from a sign problem. We contend here to study a model with a more restricted symmetry group as a representative case.

Since the four fermion term reduces the symmetry of the action at zero quark mass, the coupling $g$ does not receive additive renormalization. The chiral symmetry is fully broken by the Wilson term and restored at a critical value of the bare mass $m_0 = m_c(\beta,g)$. It is instructive to study the partially conserved axial current (PCAC) relation in the model. The relation is obtained through a variation of the action by an infinitesimal axial transformation (for details in the case of gauge theories see \cite{Luscher:1996vw}). It is identical in the diagonal and non-diagonal directions up to a term arising from the variation of the NJL term:
\begin{align}
&\partial_\mu \ev{ A^{I,d}_\mu(x) O } = 2\bar m \ev{ P^d(x) O } \label{gsigma1} \\
&- 4 a^2 \bar g^2 \left( 1- \delta^{d,3} \right ) \ev{ S^0(x) P^d(x) O },\nonumber
\end{align}
where $\bar g$ is a renormalized NJL coupling.
For convenience the order $1$ and $a$ terms have been absorbed into a renormalized axial current. Thus
\begin{align}
A_\mu^{I,d}(x) &= Z_A \bar\Psi(x) \gamma_\mu \gamma_5 \tau^d \Psi(x) + a c_A \partial_\mu P^d(x),\\
P^d(x) &= \bar \Psi(x) \gamma_5 \tau^d \Psi(x) \textrm{ and } S^0(x) = \bar\Psi(x) \Psi(x), 
\end{align}
where $\tau^d$ are the generators of the flavor symmetry group and $\tau^1,\tau^2$ and $\tau^3$ are the Pauli matrices.

The stability of the expansion is guaranteed if $\bar g$ is at most order $1$. In general we neglect any terms arising at order $a$ or higher, including the $c_A$ term, but the second term Eq.~(\ref{gsigma1}) describes the breaking of the non-diagonal axial symmetries and must be retained.

The chiral symmetry is restored when $\partial_\mu \ev{ A_\mu^{I,3} O} = 0$ and thus $\bar m=0$. On the critical surface the non-diagonal PCAC relations with $d=1$ and $2$ read
\begin{align}
&\partial_\mu \ev{ A^{I,d}_\mu(x) O } = - 4a^2 \bar g^2 \ev{ S^0(x) P^d(x) O }.  \label{pcac_nondiag}
\end{align}
The symmetry appears to be broken by an $a^2$-term at nonzero $g$. However, the scaling of the coupling $\bar g$ is non-trivial and the term does not necessarily scale as $a^2$.
The divergence of the axial current can be measured and specifically when the chiral symmetry is broken we find large values.

The non-diagonal PCAC relation provides a convenient way of measuring the chiral condensate without the need for additive renormalization.
%
The scalar density $S^0(x)$ on the right hand side of Eq.~(\ref{gsigma1}) can be split into the chiral condensate $\Sigma_L = \sum_x\ev{S^0(x)}/V$ and $S_S^0(x)=S^0(x)-\Sigma_L$.
By choosing $O= P^d(y)$ with $d=1,2$ on the critical surface, Eq.~(\ref{gsigma1}) becomes
\begin{align}
&\partial_\mu \ev{ A^{I,d}_\mu(x) P^d(y) } \label{gbar_pcac} \\ 
&=\left( 2\bar m - 4a^2 \bar g^2  \Sigma_L \right ) \ev{ P^d(x) P^d(y) }  \nonumber  \\
&- 4a^2 \bar g^2  \ev{ S_S^0(x) P^d(x) P^d(y) } . \nonumber
\end{align}
The last term vanishes at large separations and the PCAC mass $\bar m$ can be measured using the diagonal PCAC relation. The chiral condensate can then be measured by calculating the correlators in the first and second terms.

The observable $\bar g^2 \Sigma_L$ measures the breaking of the non-diagonal axial flavor symmetry, which is only broken if $\bar g$ and $\Sigma_L$ are nonzero. We may choose a renormalization scheme in which either or both receive a multiplicative renormalization. For example, it is possible to choose $\bar g = g$, in which case the renormalization coefficient for the chiral condensate depends on both couplings $g$ and $\beta$.
Here we in fact only measure the combination and do not employ separate renormalization schemes for the two quantities.

\section{Phase Diagram} \label{phasediagram}

The phase diagram of the lattice model considered here shares some features of the phase diagrams of both the lattice SU(2) adjoint model with Wilson fermions and the ungauged NJL model. In particular at strong gauge coupling, corresponding to small $\beta$s, there is a bulk phase in which the chiral zero quark line becomes a first order transition line and small quark masses can therefore not be attained.
In the weak coupling phase and zero quark mass, the phase diagram is split into two regions, an infrared conformal region at weak NJL coupling with intact chiral symmetry and a region of strong NJL coupling where the chiral symmetry is spontaneously broken.
When performing simulations on a fixed lattice volume, a region with a nonzero expectation value for the Polyakov loops is present at large enough $\beta$ corresponding to small physical volumes. At $\beta=\infty$ in this small volume region the model reduces to the ungauged NJL model studied in \cite{Rantaharju:2016jxy}.

\begin{figure} \center
\includegraphics[height=0.6\linewidth]{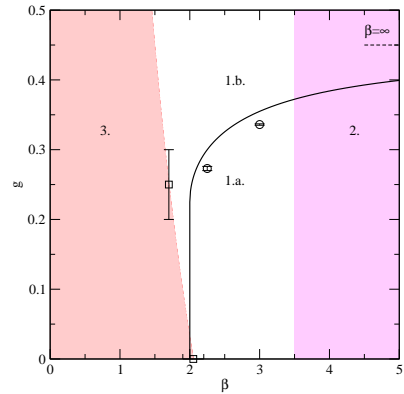}
\caption{ A sketch of the phase diagram at zero quark mass. Phases 1.a and 1.b are the physical infrared conformal and chirally broken regions. The solid line shows the large N ladder approximation result for the critical line separating the phases (the \csb line) and circles denote its measured locations.
In region 2 at $\beta>\beta_{\textrm{max}}$, the Polyakov loop grows with $\beta$, indicating significant finite size effects. At $L=16$ we find $\beta_{\textrm{max}}\gtrsim 3$. In phase 3. there is a first order bulk transition instead of a critical line and the quark mass cannot be taken to zero. The squares denote the measured boundaries of this phase. }
\label{phase_plot}
\end{figure}

A sketch of the phase diagram at zero mass is shown in Fig. \ref{phase_plot}. It includes the following significant regions:
\begin{enumerate}
\item The physically interesting region, which is split into two phases:
\begin{enumerate}[label*=\alph*.]
\item The infrared conformal phase at $g<g_c(\beta)$. The masses of all composite states approach zero on the critical surface $m=m_c(\beta,g)$ with a behavior characterized by an anomalous dimension $\gamma_m(g)$.
\item The chirally broken phase at $g>g_c(\beta)$. Here the diagonal pseudoscalar meson mass approaches zero on the critical surface as a square root of the quark mass, while other states remain massive. The combination $\bar g^2 \Sigma_L$ has a nonzero expectation value and it is as an order parameter for the broken chiral symmetry.
\end{enumerate}

\item A small volume region at $\beta>\beta_{max}(L)$. At $\beta<\beta_{max}(L)$, the Polyakov loop
\begin{align}
L_P &= \frac 14 \ev{ \sum_\mu \left| P_\mu\right | }, \\
P_\mu &= \frac{1}{L^3} \sum_{x_{\nu \neq\mu} } \prod_{x_\mu} U_{x,\mu},
\end{align}
has a small expectation value that approaches zero when $L\to\infty$. When $\beta>\beta_{max}(L)$ the expectation value grows with $\beta$, approaching $1$ at $\beta\to\infty$.
The ungauged NJL model exists in the limit $\beta\to\infty$ where the small volume region extends to the limit $L\to\infty$. The region is divided into chirally broken and symmetric phases by the $\chi SB$ line.

\item A bulk phase at small $\beta$. At $\beta<\beta_c(g)$ the critical surface is replaced by a first order transition. The transition is marked by a jump in the plaquette expectation value and, based on the observation at $g=0$, we expect a jump in the the quark mass. Since the quark mass jumps from a positive to a negative value, zero quark mass cannot be reached. 
At vanishing four-fermion coupling, it was found in previous studies that $\beta_c(0)\simeq 2$. 
Here we found that at $\beta=1.7$, the first order transition persists up to $g<0.3$.

\item An unphysical flavor-parity broken region, with a nonzero expectation value of
\begin{align}
\ev{\pi_3} = \frac 1V \ev{ \sum_x \pi_3(x) }. \label{pi3definition}
\end{align}
The critical surface in the chirally broken region 1.b is the second order transition boundary between the flavor-parity broken phase and the physical positive mass phase.
There are several unphysical phases on the negative mass side of the critical surface corresponding to different fermion doubler modes. The existence of a clear order parameter for the flavor-parity broken phase helps in identifying the critical surface without measuring the diagonal pseudoscalar meson mass.

\end{enumerate}

\begin{figure}
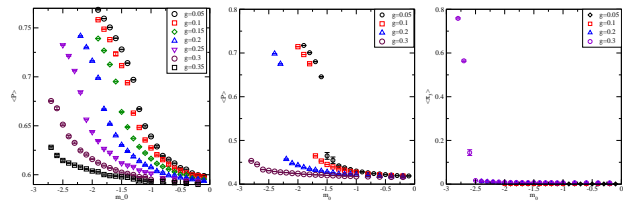
 \center
\includegraphics[height=0.3\linewidth]{{b2.25_bulk}.eps}
\includegraphics[height=0.3\linewidth]{{b1.7_bulk}.eps}
\includegraphics[height=0.3\linewidth]{{b1.7_pi}.eps}
\caption{ The plaquette expectation value as a function of the bare mass at several values of $g$ and $\beta=2.25$ (left) and $\beta=1.7$ (middle) and the expectation value $\ev{\pi_3}$ at $\beta=1.7$ (right). At the larger $\beta$ we observe only a crossover and a critical line can be found. At $\beta=1.7$ and $g<0.3$ we see a first order transition into the bulk phase. In this case the transition happens at a positive quark mass, preventing studies at the critical line. At larger $g=0.3$, we find no first order transition and there is a critical line signaled by a nonzero expectation value $\ev{\pi_3}$. }
\label{bulkphase_plot}
\end{figure}

We find numerical evidence for the phases and transitions described using lattice simulations.
We generate configurations of the gauge field $U$ and the auxiliary fields $\sigma$ and $\pi_3$ using the HMC algorithm. A full update consists of two HMC updates, a trajectory that only updates the auxiliary field, keeping the gauge field constant, and a trajectory that updates both the auxiliary fields and the gauge feld. In both cases we tune the time step to keep the acceptance rate above $80\%$. At small mass and large lattice size we use the Hasenbusch method to accelerate the HMC update \cite{Hasenbusch:2001ne}.
The scans of the parameter space are performed using the lattice size $8^4$. We produce at least 200 configurations after thermalization for each of these measurements.

The transition into the bulk phase 3 is shown in Fig.~\ref{bulkphase_plot}. In previous studies at $g=0$ \cite{Hietanen:2008mr,Catterall:2008qk} a crossover was observed at $\beta > \beta_c \approx 2$ and a first order transition at $\beta < \beta_c $. The transition can be identified by a jump in the plaquette expectation value. We observe a crossover at $\beta = 2.25$ at several values of $g$.
As the four fermion coupling $g$ is increased, the transition becomes smoother and moves to smaller bare mass. We observe what appears to be a first order transition at $\beta=1.7$ and $g<0.3$. At $\beta=1.7$ and $g=0.3$ the behavior of the plaquette again changes continuously and we find a critical mass, signified by $\ev{\pi_3}$ gaining a non-zero expectation value.

\begin{figure} \center
\includegraphics[height=0.45\linewidth]{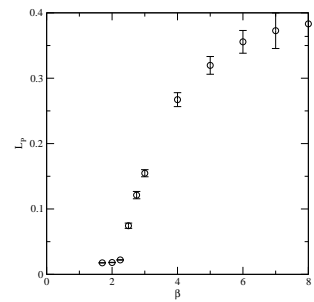}
\caption{ The Polyakov loop expectation value as a function of the gauge coupling $\beta$ at $L=8$, $g=0.1$ and $m_0=0$. }
\label{polyakov_plot}
\end{figure}

An example of the transition to the small volume region 2 is shown in Fig. \ref{polyakov_plot}. The figure shows the Polyakov loop expectation value as a function of $\beta$ at $m_0=0$. The four fermion coupling has no observable effect to the Polyakov loop in this case. The Polyakov loop is small at $\beta \lesssim 2.25$ and grows with $\beta$ when $\beta > 2.25$.
We monitor the Polyakov loop in all of our runs in order to avoid large finite size effects.

\section{The Chiral Symmetry Breaking Transition} \label{chiralbreaking}

The Nambu--Jona-Lasinio model predicts a second order transition between the chirally symmetric and broken phases \cite{Nambu:1961fr}. A strong gauge interaction also tends to cause chiral symmetry breaking and decreases the value of the critical NJL coupling needed to trigger the transition \cite{Yamawaki:1996vr}. In the more familiar case of a small number of flavors in the fundamental representation, the gauge interaction causes spontaneous chiral symmetry breaking in the full phase space of the model. A transition may nevertheless occur between a phase dominated by the gauge interaction and a phase dominated by the NJL interaction.

In the case of the SU(2) gauge interaction with 4 flavors in the fundamental representation this transition was observed in lattice studies \cite{Catterall:2011ab,Catterall:2013koa}. These results strongly point to a first order transition. In the same model a second order transition was observed in the finite volume phase.
In the adjoint SU(2) model the gauge interaction does not cause spontaneous chiral symmetry breaking and the properties of the transition may be completely different.

\begin{figure}
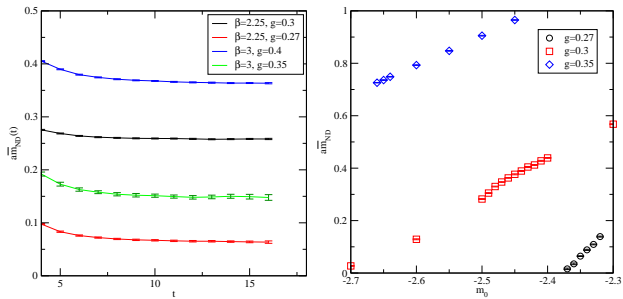
 \center
\includegraphics[height=0.45\linewidth]{ndpcacmeasurements.eps}
\includegraphics[height=0.45\linewidth]{gsigma_m.eps}
\caption{Left: The $\bar m_{ND}(t)$ as a function of $t$. The $t$ dependence of the condensate provides a measure of discretization effects. Right:  The condensate $\bar m_{ND}$ as a function of bare mass with $\beta=2.25$ and $L=16$. }
\label{gsigma_m_b2.25}
\end{figure}

We study the transition using the non-diagonal meson masses and the condensate
\begin{align}
\bar m_{ND}(t) = \frac{ \sum_{\bf x} \partial_0 \ev{ A^{ND}_\mu(t,{ \bf x}) P^{ND}(0) }}{ \sum_{\bf x } \ev{ P^{ND}(t,{\bf x}) P^{ND}(0) } }.
\end{align}
At large enough $t$, by using Eq.~(\ref{gbar_pcac}), we find
\begin{align}
\bar m_{ND}(t) = -4a^2\bar g^2 \Sigma_L.
\end{align} 
The left panel of Fig. \ref{gsigma_m_b2.25} shows several representative examples of the $t$-dependence of $\bar m_{ND}(t)$.

As noted in \cite{Rantaharju:2016jxy}, the diagonal pseudoscalar meson correlation function has a disconnected contribution. Instead of using this noisy observable, we identify the critical surface using the second order transition into the parity broken phase.
In the flavor-parity broken phase, the condensate $\ev{\pi_3}$ defined by Eq.~(\ref{pi3definition}) acquires a non-zero expectation value and the susceptibility
\begin{align}
\chi_\pi=\ev{\left(\sum_x \pi_3(x)\right)^2}-\ev{\left(\sum_x \pi_3(x)\right)}^2
\end{align}
diverges on the critical surface.

\begin{figure}
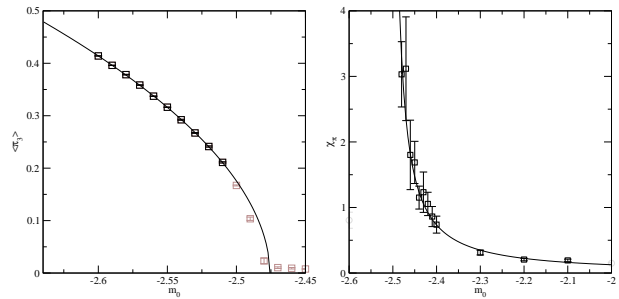
 \center
\includegraphics[height=0.45\linewidth]{{pi_ave_fit_s0.3_b2.25}.eps}
\includegraphics[height=0.45\linewidth]{{pi_susc_fit_s0.3_b2.25}.eps}
\caption{ The expectation value $\ev{\pi_3}$ (left) and the susceptibility $\chi_\pi$ (right) with $\beta=2.25$ and $g=0.3$. }
\label{gauged_pi_susc}
\end{figure}

\begin{table}[t]
\center
\begin{tabular}{ c c | c c | c c }
$L$ & $\beta$ & $g$ & $m_0$  & $m_c$      & $\chi^2/d.o.f.$ \\ \hline
16 & 2.25 & 0.27  & -2.345  & -2.348(1) & 0.89 \\
   &      & 0.275 & -2.367  & -2.3659(8) & 0.25 \\
   &      & 0.28  & -2.391  & & \\
   &      & 0.29  & -2.435  & & \\
   &      & 0.3   & -2.478  & -2.4772(9) & 0.31 \\
   &      & 0.31  & -2.518  & & \\
   &      & 0.32  & -2.558  & -2.5579(4) & 0.70  \\
   &      & 0.33  & -2.594  & & \\
   &      & 0.35  & -2.661  & -2.6609(5) & 0.73 \\
   &      & 0.4   & -2.817  & & \\
   &      & 0.5   & -3.05   & -3.0496(2) & 0.78 \\ \hline
16 & 3  & 0.35   & -2.577 & -2.577(4)  & 0.66 \\
   &    & 0.375  & -2.666 & -2.6660(5) & 0.67 \\
   &    & 0.39   & -2.715 &  &  \\
   &    & 0.4    & -2.745 & -2.745(1)  & 0.23 \\
   &    & 0.41   & -2.777 &  & \\
   &    & 0.42   & -2.806 &  & \\
   &    & 0.425  & -2.820 & -2.822(1)  & 0.82 \\
   &    & 0.43   & -2.835 &  & \\
   &    & 0.45   & -2.889 & -2.8893(8) & 0.07 \\ \hline
20 & 3  & 0.35   & -2.574 & -2.5736(3) & 0.14 \\
   &    & 0.375  & -2.665 & -2.6653(7) & 0.91 
\end{tabular}
\caption{ Simulation parameters $L$, $\beta$, $g$ and $m_0$ used to study the chiral symmetry breaking transition. The critical mass $m_c$ is recovered from the scaling fit Eq.~(\ref{pi3scaling}). The errors quoted include an estimate of systematic uncertainty found by varying the fit range. The mass $m_0$ is the actual value of the parameter used in simulations. }
\label{mcrit_table}
\end{table}

\begin{table}[t]
\center
\begin{tabular}{  c c | c c c c }
$L$ & $\beta$ & $c_0$ & $c_1$ & $c_2$ & $\chi^2/d.o.f.$ \\ \hline
16 & 2.25 & -4.220(6) & 0.679(5) & -0.047(1) & 8.8 \\
16 & 3    & -4.4(1)   & 0.86(8)  & -0.073(2) & 1.8 \\
\end{tabular}
\caption{The parametrization of the critical surface given in Eq.~(\ref{critical_line_fit}).  }
\label{table_mcrit_fit}
\end{table}

We locate the critical surface via lattice simulations at two lattice spacings, $\beta=2.25$ and $\beta=3$.  We use the lattices of size $L=16$ and, at $\beta=3$, we use two simulations at $L=20$ for comparison. Fig. \ref{gauged_pi_susc} shows $\ev{\pi_3}$ (left panel) and $\chi_\pi$ (right panel) measured at $g=0.3$ and $\beta=2.25$.
The autocorrelation times for the observables considered in this paper are at most of order 20 close to the critical regions. Our error analysis takes the autocorrelation times into account using a bootstrap blocking procedure.
As the transition is expected to be in the mean-field universality class, we fit the behavior of the condensate to
\begin{align}
\ev{\pi_3}(m_0,g) = C_\pi(g) \left( m_c(g) - m_0 \right )^{0.5} \label{pi3scaling}
\end{align}
at fixed $\beta$.
The values of $m_c(g)$ and the $\chi^2$ value per degree of freedom ($d.o.f.$) for each fit are given in Table \ref{mcrit_table}. In addition to the statistical error, we include an estimate of systematic error from varying the fit range.

We use an interpolating function to find additional parameter sets on
the critical surface, listed in table \ref{mcrit_table} without corresponding values of $m_c$. We parametrize the critical surface as
\begin{align}
m_c(\beta) = c_0(\beta) + c_1(\beta)/g + c_2(\beta)/g^2. \label{critical_line_fit}
\end{align}
The statistical and systematic errors on the critical mass are small and consequently the $\chi^2/d.o.f.$ of this fit is large. 
The values are found in table \ref{table_mcrit_fit}. 
In order to quantify the systematic error introduced by the interpolation function, we measure the condensate $\bar m_{ND}$ at three values of $g$ at $\beta=2.25$ around the transition. 
The values are shown in the right panel of Fig. \ref{gsigma_m_b2.25}. An expected deviation of $\sim 0.01$ in the mass results in a similar deviation of $\sim 0.02$ in the condensate.

\begin{table*}
\center
\begin{tabular}{ c c | c c c c c c c }
$\beta$ & $\chi^2/d.o.f.$ & $g_c$ & $C_\Sigma$ & $\beta_\Sigma$  & $C_{NDP}$ & $\beta_{NDP}$ & $C_{NDV}$ & $\beta_{NDV}$ \\ \hline
2.25 & 0.72 & 0.274(3) & 1.7(1)  & 0.52(3) & 4.1(1) & 0.33(3) & 4.1(1) & 0.31(2) \\
3    & 0.95 & 0.335(2) & 2.0(1) & 0.62(3) & 6.0(2) & 0.54(2) & 6.1(2) & 0.54(1) \\
\end{tabular}
\caption{ The scaling dimensions and coefficients in Eq.~(\ref{scaling_gsigma}) and (\ref{scaling_masses}). The errors include an estimate of the systematic error from varying the fit range. }
\label{scaling_table}
\end{table*}

\begin{figure}
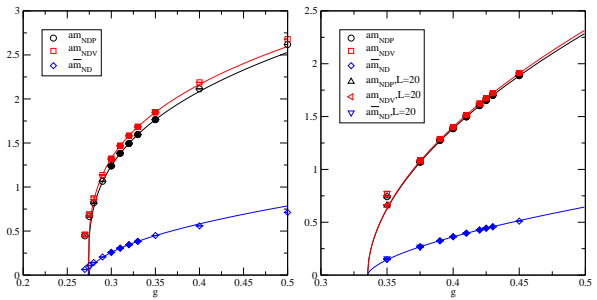
 \center
\includegraphics[height=0.45\linewidth]{{combined_scaling_b2.25}.eps}
\includegraphics[height=0.45\linewidth]{{combined_scaling_b3}.eps}
\caption{ The condensate $\bar m_{ND}$ (left) and the non-diagonal pseudoscalar ($NDP$) and vector ($NDV$) meson masses at $\beta=2.25$ (left) and $\beta=3$ (right). }
\label{ndpmass_broken_2.25}
\end{figure}

We then measure $\bar m_{ND}$ and the non-diagonal pseudoscalar ($NDP$) and non-diagonal vector ($NDV$) meson masses on the critical surface at zero quark mass in the phase 1.b, by using lattices with time extent $T=2L$.
The values of $g$ and $m_0$ used in this step are listed in table \ref{mcrit_table}. 

Fig. \ref{ndpmass_broken_2.25} shows the condensate $\bar m_{ND}$ and the masses of the non-diagonal pseudoscalar and vector mesons. 
Each of these quantities is expected to scale to zero at the second order \csb line at zero quark mass and we fit them to the lowest order behavior
\begin{align}
a\bar m_{ND} &= C_{\Sigma} \left( g-g_c \right )^{\beta_\Sigma}, \label{scaling_gsigma}\\
am_X &= C_X \left( g-g_c \right )^{\beta_X}, \label{scaling_masses}
\end{align}
where $X = NDP$ or $NDV$.
The exponents and the location of the \csb line $g_c$ are given in table \ref{scaling_table} along with the $\chi^2/d.o.f$ values. The values of $g_c$ are also shown in Fig. \ref{phase_plot}.
Systematic errors from varying the fit range are included in the values.
To control finite volume effects, at $\beta=3$ we have performed two sets of measurements at $L=20$. By comparing to $L=16$ volume, we find a significant difference at $g=0.35$ but no visible difference at $g=0.375$. 
These results indicate that the transition is compatible with a second order transition although we cannot yet rule out a weak first order transition.  
A better understanding of the order of the transition will require simulations with larger lattice sizes and closer to the critical line.

\section{Mass Anomalous Dimension} \label{anomdim}

In the infrared conformal phase 1.a chiral symmetry is not spontaneously broken: the weak four fermion interaction has no effect on the vacuum and the couplings flow to an infrared fixed point. At $g=0$ the IR fixed point has been found in previous studies with a mass anomalous dimension between $0.3$ and $0.4$. 
Here we study the IR fixed point of the model at $0<g<g_c$.  
At $g>0$ the model may be attracted to a different IR fixed point than at $g=0$ with different critical exponents. These IR fixed points would then lie on a continuous line of RG fixed points and the mass anomalous dimension would depend on the four fermion coupling $\gamma_m(g)$. 
On the other hand, if the coupling $g$ is irrelevant in the IR, the model stays in the basin of attraction of the same IR fixed point and the mass anomalous dimension must be independent of the values of $g$ up to the critical value.
Because the infrared fixed points of the renormalization group flow are stable it is expected that the critical behavior is independent of the lattice cut-off $a$.

In order to determine the value of the mass anomalous dimension, it is worth noting that the fermion matrix $D_W + m + \sigma(x) + i\gamma_5\tau_3\pi_3(x)$ is not normal and the method used in \cite{Patella:2012da} fails. 
Here we use a direct method of deforming the IR conformal model with a fermion mass. 
In this case, the masses of all states follow the hyperscaling relation 
\begin{align}
  &Lm_X  = f(x) \label{eq_fss}
\end{align}
where $x\equiv L \left | m_0-m_c \right |^\frac{1}{1+\gamma_m}$.
We measure the masses of the non-diagonal pseudoscalar and vector mesons using four lattice sizes, $L=16$, $18$, $20$ and $24$ with lattice dimensions $L^3\times2L$, at $\beta=2.25$, and several values of the bare mass.

\begin{table}[t]
\center
\begin{tabular}{ c c | c c }
  $g$ & $\chi^2/d.o.f.$ & $m_c$ & $\gamma_m$ \\ \hline
  $0.05$ & 1.5  & $-1.241(3)^{+0.005}_{-0.005}$  & $0.5(1)^{+0.2}_{-0.1}$ \\
  $0.1$  & 0.6  & $-1.357(1)^{+0.001}_{-0.004}$  & $0.54(6)^{+0.2}_{-0.2}$   \\
  $0.2$  & 0.9  & $-1.8276(5)^{+0.001}_{-0.001}$ & $0.89(3)^{+0.1}_{-0.04}$   \\
  $0.25$ & 1.0  & $-2.196(1)^{+0.001}_{-0.002}$ & $1.06(5)^{+0.1}_{-0.05}$
\end{tabular}
\caption{ Results from hyperscaling fits. The second error estimates the effect of varying the fit range.  }
\label{hyperscalingtable}
\end{table}

\begin{figure}
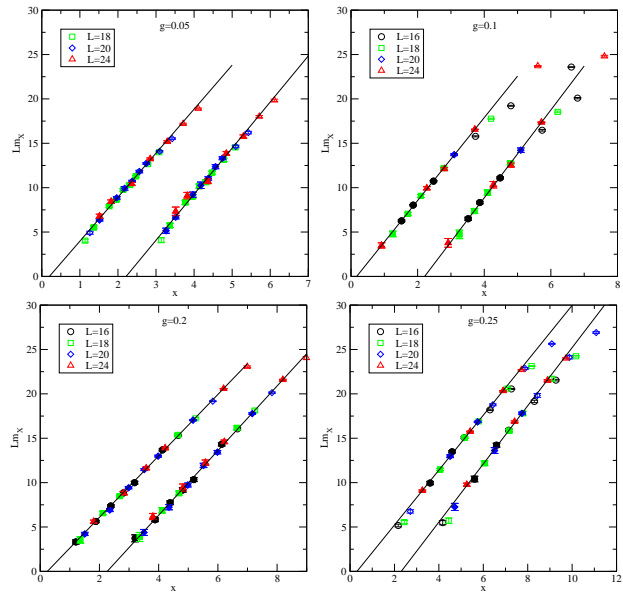
 \center
\includegraphics[height=0.45\linewidth]{{nd_combined_scaling_fit_s0.05}.eps}
\includegraphics[height=0.45\linewidth]{{nd_combined_scaling_fit_s0.1}.eps}
\includegraphics[height=0.45\linewidth]{{nd_combined_scaling_fit_s0.2}.eps}
\includegraphics[height=0.45\linewidth]{{nd_combined_scaling_fit_s0.25}.eps}
\caption{ A hyperscaling fit to the non-diagonal meson masses at $g=0.05$, $0.1$, $0.2$ and $0.25$ and $\beta=2.25$. The data points with filled in symbols are included in the fit. The points corresponding to the vector meson have been shifted to the right by $\Delta x=2$.}
\label{b2.25_hyperscaling}
\end{figure}

In order to estimate $\gamma_m$, we make use of the asymptotic behavior of $f(x)$ as $x\to \infty$:
\begin{align}
  &f(x) = a_X x + b_X. \label{eq_fss_asy}
\end{align}
The linear behavior is only valid for large enough $x$, provided that Eq.~(\ref{eq_fss}) applies, i.e. that $L$ is large enough and $m_0-m_c$ is sufficiently small. 

As the exact range of validity of the asymptotic finite size scaling formula above is not known \textit{a priori}, we perform a fit to Eq.~(\ref{eq_fss_asy}) and exclude the data points at heavy quark masses and small $x$ so that the final fit describes the data well. 
The fits are shown in Fig. \ref{b2.25_hyperscaling} for four different values of the coupling $g$ and the values of $m_c$ and $\gamma_m$ so obtained are reported in table \ref{hyperscalingtable}.
The measurements have relatively large systematic errors from the variation of the fit range, but 
the fit is fairly robust in all cases.

At small NJL coupling, $g=0.05$ and $g=0.1$ we find an anomalous dimension $\gamma_m=0.5(2)$, compatible with estimates at $g=0$. However, when we increase the NJL coupling to $g=0.2$ and $g=0.25$, we find an increasing anomalous dimension. Interestingly, the anomalous dimension is compatible with $1$ close to the $\chi SB$ line.

In addition to the possibility of a line of fixed points, there is another possible interpretation of the result. The two lower values of $g$ could be attracted to the IRFP at $g=0$ and have the same anomalous dimension. Similarly, the anomalous dimensions at the two larger values of the NJL coupling are compatible with each other. The model at these two parameter values could be attracted to a single second fixed point. The basins of attraction of the two fixed points would form two distinct phases within region 1.a. However we observe no evidence for a phase transition inside the infrared conformal region.

\section{Conclusions} \label{conclusions}

We provided a first mapping of the phase diagram of the $SU(2)$ gauged NJL model with two flavors of fermions in the adjoint representation regularized on a lattice. This first numerical evidence supports the conjectures put forward in \cite{Fukano:2010yv,Yamawaki:1996vr}. The phase diagram presents some features observed in the phase diagrams of an infrared conformal gauge model and a pure NJL model.
In particular, in the weak coupling phase of the lattice model, there is an infrared conformal phase at small $g$ and a transition into a chirally broken phase at a critical value $g_c$ of the four fermion coupling. 

In the ungauged NJL model the transition into the chirally broken phase is second order. The addition of a gauge interaction can modify the dynamics of the transition and in \cite{Catterall:2011ab,Catterall:2013koa} a first order transition was found in a model with no infrared conformal phase.
Here we studied the transition with two different gauge couplings and found a behavior compatible with a second order transition. 
However, larger lattices will be required to reach deeper into the critical region of small masses and, with the present data, a first order transition \cite{Sannino:2012wy} cannot be ruled out. 

The mass anomalous dimension $\gamma_m$ has been investigated, which is an important quantity for model building beyond the Standard Model. 
In various approximations it has a large value close to the $\chi SB$ line. 
In Ideal Walking, the NJL interaction is used to break chiral symmetry, naturally creating a walking model with a large mass anomalous dimension. 
We measured $\gamma_m(g)$ at four values of the four fermion coupling $g$ and found larger values with increasing NJL coupling. 
Close to the \csb line we find $\gamma_m\sim1$. The systematic errors on the values of $\gamma_m$ are relatively large and we plan to improve them by measuring the anomalous dimension with additional values of the NJL coupling and using larger lattice sizes at smaller quark masses. 
This preliminary analysis, still needing further tests, indicates that the present model can be viewed as the first realization of the Ideal Walking scenario.

In the future it would therefore be very interesting to investigate the low energy spectrum of the theory near the Ideal Walking region, including the mass of the spin-one resonances and the lightest scalar state to see if the former are near degenerate \cite{Appelquist:1998xf} and the latter displays pseudo-dilaton couplings \cite{Sannino:1999qe,Dietrich:2005jn,Dietrich:2006cm,Goldberger:2008zz,Matsuzaki:2013eva,Golterman:2016lsd,Crewther:2012wd,Crewther:2013vea} and a parametrically light mass \cite{Dietrich:2005jn}. 
To this end one can compare the spectrum and couplings with the isosinglet scalar extended chiral perturbation theory of \cite{Hansen:2016fri,Appelquist:2017wcg}. 
Finally it would be interesting to test intriguing holographic descriptions of gNJL models \cite{Clemens:2017udk} with our realization.

\section{Acknowledgments}

This work was supported by the Danish National Research Foundation DNRF:90 grant, by a Lundbeck Foundation Fellowship grant and the U.S. Department of Energy, Office of Science, Nuclear Physics program under Award Number DE-FG02-05ER41368. Computing facilities were provided by the Danish Center for Scientific Computing and the DeIC national HPC center at SDU and the Extreme Science and Engineering Discovery Environment (XSEDE), which is supported by National Science Foundation grant number ACI-1053575.


\begin{thebibliography}{9999}

\bibitem{Weinberg:1975gm}
  S.~Weinberg,
  Phys.\ Rev.\ D {\bf 13} (1976) 974.
  doi:10.1103/PhysRevD.13.974


\bibitem{Susskind:1978ms}
  L.~Susskind,
  Phys.\ Rev.\ D {\bf 20} (1979) 2619.
  doi:10.1103/PhysRevD.20.2619


\bibitem{Kaplan:1983fs}
  D.~B.~Kaplan and H.~Georgi,
  Phys.\ Lett.\  {\bf 136B} (1984) 183.
  doi:10.1016/0370-2693(84)91177-8


\bibitem{Kaplan:1983sm}
  D.~B.~Kaplan, H.~Georgi and S.~Dimopoulos,
  Phys.\ Lett.\  {\bf 136B} (1984) 187.
  doi:10.1016/0370-2693(84)91178-X


\bibitem{Cacciapaglia:2015yra}
  G.~Cacciapaglia and F.~Sannino,
  Phys.\ Lett.\ B {\bf 755} (2016) 328
  doi:10.1016/j.physletb.2016.02.034
  [arXiv:1508.00016 [hep-ph]].


\bibitem{Sannino:2016sfx}
  F.~Sannino, A.~Strumia, A.~Tesi and E.~Vigiani,
  JHEP {\bf 1611} (2016) 029
  doi:10.1007/JHEP11(2016)029
  [arXiv:1607.01659 [hep-ph]].


\bibitem{Pica:2016rmv}
  C.~Pica and F.~Sannino,
  Phys.\ Rev.\ D {\bf 94} (2016) no.7,  071702
  doi:10.1103/PhysRevD.94.071702
  [arXiv:1604.02572 [hep-ph]].


\bibitem{Ferretti:2016upr}
  G.~Ferretti,
  JHEP {\bf 1606} (2016) 107
  doi:10.1007/JHEP06(2016)107
  [arXiv:1604.06467 [hep-ph]].


\bibitem{Fukano:2010yv}
  H.~S.~Fukano and F.~Sannino,
  Phys.\ Rev.\ D {\bf 82} (2010) 035021
  doi:10.1103/PhysRevD.82.035021
  [arXiv:1005.3340 [hep-ph]].


\bibitem{Holdom:1981rm}
  B.~Holdom,
  Phys.\ Rev.\ D {\bf 24} (1981) 1441.
  doi:10.1103/PhysRevD.24.1441


\bibitem{Appelquist:1988fm}
  T.~Appelquist, M.~Soldate, T.~Takeuchi and L.~C.~R.~Wijewardhana,
  YCTP-P19-88.


\bibitem{Kondo:1988qd}
  K.~i.~Kondo, H.~Mino and K.~Yamawaki,
  Phys.\ Rev.\ D {\bf 39} (1989) 2430.
  doi:10.1103/PhysRevD.39.2430


\bibitem{Holdom:1995fu}
  B.~Holdom,
  Phys.\ Rev.\ D {\bf 54} (1996) 1068
  doi:10.1103/PhysRevD.54.1068
  [hep-ph/9512298].


\bibitem{Sannino:2010ca}
  F.~Sannino,
  Phys.\ Rev.\ D {\bf 82} (2010) 081701
  doi:10.1103/PhysRevD.82.081701
  [arXiv:1006.0207 [hep-lat]].


\bibitem{Pica:2010mt}
  C.~Pica and F.~Sannino,
  Phys.\ Rev.\ D {\bf 83} (2011) 116001
  doi:10.1103/PhysRevD.83.116001
  [arXiv:1011.3832 [hep-ph]].


\bibitem{Pica:2010xq}
  C.~Pica and F.~Sannino,
  Phys.\ Rev.\ D {\bf 83} (2011) 035013
  doi:10.1103/PhysRevD.83.035013
  [arXiv:1011.5917 [hep-ph]].


\bibitem{Ryttov:2010iz}
  T.~A.~Ryttov and R.~Shrock,
  Phys.\ Rev.\ D {\bf 83} (2011) 056011
  doi:10.1103/PhysRevD.83.056011
  [arXiv:1011.4542 [hep-ph]].


\bibitem{Ryttov:2016ner}
  T.~A.~Ryttov and R.~Shrock,
  Phys.\ Rev.\ D {\bf 94} (2016) no.10,  105015
  doi:10.1103/PhysRevD.94.105015
  [arXiv:1607.06866 [hep-th]].


\bibitem{Pica:2017gcb}
  C.~Pica,
  PoS LATTICE {\bf 2016} (2016) 015
  [arXiv:1701.07782 [hep-lat]].


\bibitem{Sannino:2004qp}
  F.~Sannino and K.~Tuominen,
  Phys.\ Rev.\ D {\bf 71} (2005) 051901
  doi:10.1103/PhysRevD.71.051901
  [hep-ph/0405209].


\bibitem{Dietrich:2006cm}
  D.~D.~Dietrich and F.~Sannino,
  Phys.\ Rev.\ D {\bf 75} (2007) 085018
  doi:10.1103/PhysRevD.75.085018
  [hep-ph/0611341].


\bibitem{Sannino:2012wy}
  F.~Sannino,
  Mod.\ Phys.\ Lett.\ A {\bf 28} (2013) 1350127
  doi:10.1142/S0217732313501277
  [arXiv:1205.4246 [hep-ph]].


\bibitem{Yamawaki:1996vr}
  K.~Yamawaki,
  hep-ph/9603293.


\bibitem{Fitzpatrick:2007sa}
  A.~L.~Fitzpatrick, G.~Perez and L.~Randall,
  Phys.\ Rev.\ Lett.\  {\bf 100} (2008) 171604
  doi:10.1103/PhysRevLett.100.171604
  [arXiv:0710.1869 [hep-ph]].


\bibitem{Albrecht:2009xr}
  M.~E.~Albrecht, M.~Blanke, A.~J.~Buras, B.~Duling and K.~Gemmler,
  JHEP {\bf 0909} (2009) 064
  doi:10.1088/1126-6708/2009/09/064
  [arXiv:0903.2415 [hep-ph]].


\bibitem{Parolini:2014rza}
  A.~Parolini,
  Phys.\ Rev.\ D {\bf 90} (2014) no.11,  115026
  doi:10.1103/PhysRevD.90.115026
  [arXiv:1405.4875 [hep-ph]].


\bibitem{Rattazzi:2008pe}
  R.~Rattazzi, V.~S.~Rychkov, E.~Tonni and A.~Vichi,
  JHEP {\bf 0812} (2008) 031
  doi:10.1088/1126-6708/2008/12/031
  [arXiv:0807.0004 [hep-th]].


\bibitem{Rychkov:2009ij}
  V.~S.~Rychkov and A.~Vichi,
  Phys.\ Rev.\ D {\bf 80} (2009) 045006
  doi:10.1103/PhysRevD.80.045006
  [arXiv:0905.2211 [hep-th]].


\bibitem{Antipin:2014mga}
  O.~Antipin, E.~Mølgaard and F.~Sannino,
  JHEP {\bf 1506} (2015) 030
  doi:10.1007/JHEP06(2015)030
  [arXiv:1406.6166 [hep-th]].

\bibitem{Nambu:1961fr}
  Y.~Nambu and G.~Jona-Lasinio,
  Phys.\ Rev.\  {\bf 124} (1961) 246.
  doi:10.1103/PhysRev.124.246



\bibitem{Dietrich:2005jn}
  D.~D.~Dietrich, F.~Sannino and K.~Tuominen,
  Phys.\ Rev.\ D {\bf 72} (2005) 055001
  doi:10.1103/PhysRevD.72.055001
  [hep-ph/0505059].


\bibitem{Catterall:2007yx}
  S.~Catterall and F.~Sannino,
  Phys.\ Rev.\ D {\bf 76} (2007) 034504
  doi:10.1103/PhysRevD.76.034504
  [arXiv:0705.1664 [hep-lat]].


\bibitem{Hietanen:2008mr}
  A.~J.~Hietanen, J.~Rantaharju, K.~Rummukainen and K.~Tuominen,
  JHEP {\bf 0905} (2009) 025
  doi:10.1088/1126-6708/2009/05/025
  [arXiv:0812.1467 [hep-lat]].


\bibitem{DelDebbio:2008zf}
  L.~Del Debbio, A.~Patella and C.~Pica,
  Phys.\ Rev.\ D {\bf 81} (2010) 094503
  doi:10.1103/PhysRevD.81.094503
  [arXiv:0805.2058 [hep-lat]].


\bibitem{Catterall:2008qk}
  S.~Catterall, J.~Giedt, F.~Sannino and J.~Schneible,
  JHEP {\bf 0811} (2008) 009
  doi:10.1088/1126-6708/2008/11/009
  [arXiv:0807.0792 [hep-lat]].


\bibitem{Bursa:2009we}
  F.~Bursa, L.~Del Debbio, L.~Keegan, C.~Pica and T.~Pickup,
  Phys.\ Rev.\ D {\bf 81} (2010) 014505
  doi:10.1103/PhysRevD.81.014505
  [arXiv:0910.4535 [hep-ph]].


\bibitem{DelDebbio:2009fd}
  L.~Del Debbio, B.~Lucini, A.~Patella, C.~Pica and A.~Rago,
  Phys.\ Rev.\ D {\bf 80} (2009) 074507
  doi:10.1103/PhysRevD.80.074507
  [arXiv:0907.3896 [hep-lat]].


\bibitem{DeGrand:2011qd}
  T.~DeGrand, Y.~Shamir and B.~Svetitsky,
  Phys.\ Rev.\ D {\bf 83} (2011) 074507
  doi:10.1103/PhysRevD.83.074507
  [arXiv:1102.2843 [hep-lat]].


\bibitem{DelDebbio:2010hx}
  L.~Del Debbio, B.~Lucini, A.~Patella, C.~Pica and A.~Rago,
  Phys.\ Rev.\ D {\bf 82} (2010) 014510
  doi:10.1103/PhysRevD.82.014510
  [arXiv:1004.3206 [hep-lat]].


\bibitem{DelDebbio:2010hu}
  L.~Del Debbio, B.~Lucini, A.~Patella, C.~Pica and A.~Rago,
  Phys.\ Rev.\ D {\bf 82} (2010) 014509
  doi:10.1103/PhysRevD.82.014509
  [arXiv:1004.3197 [hep-lat]].


\bibitem{Patella:2012da}
  A.~Patella,
  Phys.\ Rev.\ D {\bf 86} (2012) 025006
  doi:10.1103/PhysRevD.86.025006
  [arXiv:1204.4432 [hep-lat]].


\bibitem{Giedt:2012rj}
  J.~Giedt and E.~Weinberg,
  Phys.\ Rev.\ D {\bf 85} (2012) 097503
  doi:10.1103/PhysRevD.85.097503
  [arXiv:1201.6262 [hep-lat]].


\bibitem{Bergner:2015jdn}
  G.~Bergner, P.~Giudice, I.~Montvay, G.~Münster and S.~Piemonte,
  PoS LATTICE {\bf 2015} (2016) 227
  [arXiv:1511.05097 [hep-lat]].


\bibitem{Rantaharju:2015yva}
  J.~Rantaharju, T.~Rantalaiho, K.~Rummukainen and K.~Tuominen,
  Phys.\ Rev.\ D {\bf 93} (2016) no.9,  094509
  doi:10.1103/PhysRevD.93.094509
  [arXiv:1510.03335 [hep-lat]].


\bibitem{DelDebbio:2015byq}
  L.~Del Debbio, B.~Lucini, A.~Patella, C.~Pica and A.~Rago,
  Phys.\ Rev.\ D {\bf 93} (2016) no.5,  054505
  doi:10.1103/PhysRevD.93.054505
  [arXiv:1512.08242 [hep-lat]].


\bibitem{Rantaharju:2015cne}
  J.~Rantaharju,
  Phys.\ Rev.\ D {\bf 93} (2016) no.9,  094516
  doi:10.1103/PhysRevD.93.094516
  [arXiv:1512.02793 [hep-lat]].


\bibitem{Bergner:2016hip}
  G.~Bergner, P.~Giudice, I.~Montvay, G.~Münster and S.~Piemonte,
  arXiv:1610.01576 [hep-lat].


\bibitem{Catterall:2011ab}
  S.~Catterall, R.~Galvez, J.~Hubisz, D.~Mehta and A.~Veernala,
  Phys.\ Rev.\ D {\bf 86} (2012) 034502
  doi:10.1103/PhysRevD.86.034502
  [arXiv:1112.1855 [hep-lat]].


\bibitem{Catterall:2013koa}
  S.~Catterall and A.~Veernala,
  Phys.\ Rev.\ D {\bf 87} (2013) no.11,  114507
  doi:10.1103/PhysRevD.87.114507
  [arXiv:1303.6187 [hep-lat]].


\bibitem{Krog:2015bca}
  J.~Krog, M.~Mojaza and F.~Sannino,
  Phys.\ Rev.\ D {\bf 92} (2015) no.8,  085043
  doi:10.1103/PhysRevD.92.085043
  [arXiv:1506.02642 [hep-ph]].


\bibitem{Bardeen:1989ds}
  W.~A.~Bardeen, C.~T.~Hill and M.~Lindner,
  Phys.\ Rev.\ D {\bf 41} (1990) 1647.
  doi:10.1103/PhysRevD.41.1647


\bibitem{Chivukula:1992pm}
  R.~S.~Chivukula, M.~Golden and E.~H.~Simmons,
  Phys.\ Rev.\ Lett.\  {\bf 70} (1993) 1587
  doi:10.1103/PhysRevLett.70.1587
  [hep-ph/9210276].


\bibitem{Bardeen:1993pj}
  W.~A.~Bardeen, C.~T.~Hill and D.~U.~Jungnickel,
  Phys.\ Rev.\ D {\bf 49} (1994) 1437
  doi:10.1103/PhysRevD.49.1437
  [hep-th/9307193].


\bibitem{Kondo:1992sq}
  K.~i.~Kondo, M.~Tanabashi and K.~Yamawaki,
  Prog.\ Theor.\ Phys.\  {\bf 89} (1993) 1249
  doi:10.1143/ptp/89.6.1249, 10.1143/PTP.89.1249
  [hep-ph/9212208].


\bibitem{Harada:1994wy}
  M.~Harada, Y.~Kikukawa, T.~Kugo and H.~Nakano,
  Prog.\ Theor.\ Phys.\  {\bf 92} (1994) 1161
  doi:10.1143/PTP.92.1161, 10.1143/ptp/92.6.1161
  [hep-ph/9407398].


\bibitem{Kubota:1999jf}
  K.~I.~Kubota and H.~Terao,
  Prog.\ Theor.\ Phys.\  {\bf 102} (1999) 1163
  doi:10.1143/PTP.102.1163
  [hep-th/9908062].


\bibitem{Litim:2014uca}
  D.~F.~Litim and F.~Sannino,
  JHEP {\bf 1412} (2014) 178
  doi:10.1007/JHEP12(2014)178
  [arXiv:1406.2337 [hep-th]].


\bibitem{Litim:2015iea}
  D.~F.~Litim, M.~Mojaza and F.~Sannino,
  JHEP {\bf 1601} (2016) 081
  doi:10.1007/JHEP01(2016)081
  [arXiv:1501.03061 [hep-th]].


\bibitem{Intriligator:2015xxa}
  K.~Intriligator and F.~Sannino,
  JHEP {\bf 1511} (2015) 023
  doi:10.1007/JHEP11(2015)023
  [arXiv:1508.07411 [hep-th]].


\bibitem{Bajc:2016efj}
  B.~Bajc and F.~Sannino,
  JHEP {\bf 1612} (2016) 141
  doi:10.1007/JHEP12(2016)141
  [arXiv:1610.09681 [hep-th]].


\bibitem{Abel:2017ujy} 
  S.~Abel and F.~Sannino,
  arXiv:1704.00700 [hep-ph].

\bibitem{Channuie:2011rq}
  P.~Channuie, J.~J.~Joergensen and F.~Sannino,
  JCAP {\bf 1105} (2011) 007
  doi:10.1088/1475-7516/2011/05/007
  [arXiv:1102.2898 [hep-ph]].


\bibitem{Bezrukov:2011mv}
  F.~Bezrukov, P.~Channuie, J.~J.~Joergensen and F.~Sannino,
  Phys.\ Rev.\ D {\bf 86} (2012) 063513
  doi:10.1103/PhysRevD.86.063513
  [arXiv:1112.4054 [hep-ph]].


\bibitem{Anguelova:2014dza}
  L.~Anguelova, P.~Suranyi and L.~C.~R.~Wijewardhana,
  Nucl.\ Phys.\ B {\bf 899} (2015) 651
  doi:10.1016/j.nuclphysb.2015.08.020
  [arXiv:1412.8422 [hep-th]].


\bibitem{Anguelova:2015dgt}
  L.~Anguelova,
  Nucl.\ Phys.\ B {\bf 911} (2016) 480
  doi:10.1016/j.nuclphysb.2016.08.020
  [arXiv:1512.08556 [hep-th]].


\bibitem{Channuie:2016iyy}
  P.~Channuie and C.~Xiong,
  Phys.\ Rev.\ D {\bf 95} (2017) no.4,  043521
  doi:10.1103/PhysRevD.95.043521
  [arXiv:1609.04698 [hep-ph]].


\bibitem{Inagaki:2016vkf}
  T.~Inagaki, S.~D.~Odintsov and H.~Sakamoto,
  Nucl.\ Phys.\ B {\bf 919} (2017) 297
  doi:10.1016/j.nuclphysb.2017.03.024
  [arXiv:1611.00210 [hep-ph]].


\bibitem{Luscher:1996vw}
  M.~Luscher and P.~Weisz,
  Nucl.\ Phys.\ B {\bf 479} (1996) 429
  doi:10.1016/0550-3213(96)00448-8
  [hep-lat/9606016].


\bibitem{Rantaharju:2016jxy}
  J.~Rantaharju, V.~Drach, C.~Pica and F.~Sannino,
  Phys.\ Rev.\ D {\bf 95} (2017) no.1,  014508
  doi:10.1103/PhysRevD.95.014508
  [arXiv:1609.08051 [hep-lat]].


\bibitem{Hasenbusch:2001ne}
  M.~Hasenbusch,
  Phys.\ Lett.\ B {\bf 519} (2001) 177
  doi:10.1016/S0370-2693(01)01102-9
  [hep-lat/0107019].


\bibitem{Appelquist:1998xf}
  T.~Appelquist and F.~Sannino,
  Phys.\ Rev.\ D {\bf 59} (1999) 067702
  doi:10.1103/PhysRevD.59.067702
  [hep-ph/9806409].


\bibitem{Sannino:1999qe}
  F.~Sannino and J.~Schechter,
  Phys.\ Rev.\ D {\bf 60} (1999) 056004
  doi:10.1103/PhysRevD.60.056004
  [hep-ph/9903359].


\bibitem{Goldberger:2008zz}
  W.~D.~Goldberger, B.~Grinstein and W.~Skiba,
  Phys.\ Rev.\ Lett.\  {\bf 100} (2008) 111802
  doi:10.1103/PhysRevLett.100.111802
  [arXiv:0708.1463 [hep-ph]].


\bibitem{Matsuzaki:2013eva}
  S.~Matsuzaki and K.~Yamawaki,
  Phys.\ Rev.\ Lett.\  {\bf 113} (2014) no.8,  082002
  doi:10.1103/PhysRevLett.113.082002
  [arXiv:1311.3784 [hep-lat]].


\bibitem{Golterman:2016lsd}
  M.~Golterman and Y.~Shamir,
  Phys.\ Rev.\ D {\bf 94} (2016) no.5,  054502
  doi:10.1103/PhysRevD.94.054502
  [arXiv:1603.04575 [hep-ph]].


\bibitem{Crewther:2012wd}
  R.~J.~Crewther and L.~C.~Tunstall,
  arXiv:1203.1321 [hep-ph].


\bibitem{Crewther:2013vea}
  R.~J.~Crewther and L.~C.~Tunstall,
  Phys.\ Rev.\ D {\bf 91} (2015) no.3,  034016
  doi:10.1103/PhysRevD.91.034016
  [arXiv:1312.3319 [hep-ph]].


\bibitem{Hansen:2016fri}
  M.~Hansen, K.~Langæble and F.~Sannino,
  Phys.\ Rev.\ D {\bf 95} (2017) no.3,  036005
  doi:10.1103/PhysRevD.95.036005
  [arXiv:1610.02904 [hep-ph]].


\bibitem{Appelquist:2017wcg}
  T.~Appelquist, J.~Ingoldby and M.~Piai,
  arXiv:1702.04410 [hep-ph].


\bibitem{Clemens:2017udk}
  W.~Clemens and N.~Evans,
  arXiv:1702.08693 [hep-th].

\end{thebibliography}
\end{document}